\documentstyle[psfig]{mn}

\newif\ifAMStwofonts

\def\simlt{\lower.5ex\hbox{$\; \buildrel < \over \sim \;$}}
\def\simgt{\lower.5ex\hbox{$\; \buildrel > \over \sim \;$}}

\ifoldfss
  \ifCUPmtlplainloaded \else
    \NewTextAlphabet{textbfit} {cmbxti10} {}
    \NewTextAlphabet{textbfss} {cmssbx10} {}
    \NewMathAlphabet{mathbfit} {cmbxti10} {} 
    \NewMathAlphabet{mathbfss} {cmssbx10} {} 
  \fi
  \ifAMStwofonts
    \ifCUPmtlplainloaded \else
      \NewSymbolFont{upmath} {eurm10}
      \NewSymbolFont{AMSa} {msam10}
      \NewMathSymbol{\upi}     {0}{upmath}{19}
      \NewMathSymbol{\umu}     {0}{upmath}{16}
      \NewMathSymbol{\upartial}{0}{upmath}{40}
      \NewMathSymbol{\leqslant}{3}{AMSa}{36}
      \NewMathSymbol{\geqslant}{3}{AMSa}{3E}

    \fi
  \fi
\fi 

\ifnfssone
  \newmathalphabet{\mathit}
  \addtoversion{normal}{\mathit}{cmr}{m}{it}
  \addtoversion{bold}{\mathit}{cmr}{bx}{it}
  \newmathalphabet{\mathbfit} 
  \addtoversion{normal}{\mathbfit}{cmr}{bx}{it}
  \addtoversion{bold}{\mathbfit}{cmr}{bx}{it}
  \newmathalphabet{\mathbfss} 
  \addtoversion{normal}{\mathbfss}{cmss}{bx}{n}
  \addtoversion{bold}{\mathbfss}{cmss}{bx}{n}
  \ifAMStwofonts
    \ifCUPmtlplainloaded \else
      \UseAMStwoboldmath
      \makeatletter
      \new@mathgroup\upmath@group
      \define@mathgroup\mv@normal\upmath@group{eur}{m}{n}
      \define@mathgroup\mv@bold\upmath@group{eur}{b}{n}
      \edef\UPM{\hexnumber\upmath@group}
      \new@mathgroup\amsa@group
      \define@mathgroup\mv@normal\amsa@group{msa}{m}{n}
      \define@mathgroup\mv@bold\amsa@group{msa}{m}{n}
      \edef\AMSa{\hexnumber\amsa@group}
      \makeatother
      \mathchardef\upi="0\UPM19
      \mathchardef\umu="0\UPM16
      \mathchardef\upartial="0\UPM40
      \mathchardef\leqslant="3\AMSa36
      \mathchardef\geqslant="3\AMSa3E
    \fi
  \fi
\fi 

\ifnfsstwo
  \DeclareMathAlphabet{\mathbfit}{OT1}{cmr}{bx}{it}
  \SetMathAlphabet\mathbfit{bold}{OT1}{cmr}{bx}{it}
  \DeclareMathAlphabet{\mathbfss}{OT1}{cmss}{bx}{n}
  \SetMathAlphabet\mathbfss{bold}{OT1}{cmss}{bx}{n}
  \ifAMStwofonts
    \ifCUPmtlplainloaded \else
      \DeclareSymbolFont{UPM}{U}{eur}{m}{n}
      \SetSymbolFont{UPM}{bold}{U}{eur}{b}{n}
      \DeclareSymbolFont{AMSa}{U}{msa}{m}{n}
      \DeclareMathSymbol{\upi}{0}{UPM}{"19}
      \DeclareMathSymbol{\umu}{0}{UPM}{"16}
      \DeclareMathSymbol{\upartial}{0}{UPM}{"40}
      \DeclareMathSymbol{\leqslant}{3}{AMSa}{"36}
      \DeclareMathSymbol{\geqslant}{3}{AMSa}{"3E}
    \fi
  \fi
\fi 

\ifCUPmtlplainloaded \else
  \ifAMStwofonts \else 
    \def\upi{\pi}
    \def\umu{\mu}
    \def\upartial{\partial}
  \fi
\fi

\title[Helioseismology and Dark Matter]
    {Helioseismology as a New Constraint on SUSY Dark Matter}
\author[Lopes, Silk \& Hansen]
{Il\'\i dio P. Lopes, Joseph Silk and Steen H. Hansen\\
Nuclear \& Astrophysics Lab., 1 Keble Road, Oxford OX1 3RH, United
Kingdom}
\date{Draft version \today}
\pubyear{2001}

\begin{document}

\maketitle


\begin{abstract}
The presence of dark matter in the solar neighbourhood can be tested
in the framework of stellar evolution theory by using the new results
of helioseismology. If weakly interacting massive particles accumulate
in the center of the Sun, they can provide an additional mechanism for
transferring energy from the solar core.  The presence of these
particles produces a change in the local luminosity of the Sun of the
order of $0.1 \%$, which is now within the reach of seismic solar
experiments.  We find that typical WIMPs with a mass of 60 GeV,
annihilation cross-section of $10^{-32} \;
\mbox{cm}^3/\mbox{sec}$, and scalar scattering interaction between
$10^{-36}\;\mbox{cm}^2$ and $10^{-41}\; \mbox{cm}^2$, the lower end of
this range being that suggested by the DAMA direct detection
experiment, modify the radial profile of the square of the sound speed
in the solar core by  $\sim 0.1 \%$. This level of change is strongly
constrained by the most recent helioseismological
results. Significantly more massive WIMPs have a much smaller effect
on the solar core temperature profile.  However future
helioseismological experiments have the potential exploring most, if
not all, of the WIMP parameter space accessible to  current
and planned direct
detection experiments.
\end{abstract}



\section{Introduction}

The dynamical behaviour of various astronomical objects, from galaxies
to galaxy clusters and to large-scale structure, in the observed
universe can only be understood if the dominant component of the mean
matter density is dark, amounting to $\Omega_m=0.3\pm 0.1.$
Constraints from primordial nucleosynthesis of the light elements
provide a compelling measure of the mean baryon abundance,
$\Omega_b=0.05\pm 0.02.$ The bulk of the dark matter is consequently
non-baryonic, and the existence of particles that interact with
ordinary matter on the scale of the weak force, so-called Weakly
Interacting Massive Particles (WIMPs), provides one of the
best-motivated candidates, arising from the lightest stable particle
predicted by  super-symmetry theories (SUSY), for resolving this problem.

One can show (Lee \& Weinberg 1977) that if such a stable particle
exists, its relic abundance is $\Omega_x h^{2} \simeq 3\times 10^{-27}
cm^3 s^{-1}/ \langle\sigma_a v\rangle$, where $\langle\sigma_a
v\rangle$ is the thermally averaged product of annihilation
cross-section and relative velocity.  Alternatively, if a new particle
with weak-scale interactions exists, then its annihilation
cross-section can be estimated to be $\langle\sigma_a v\rangle\sim
10^{-25} cm^3 s^{-1} $ (Jungman {\it et.al.} 1996).  Indeed, this
constitutes one of the strongest arguments for considering the WIMP to
be one of the best particle candidates for the non-baryonic component
of the dark matter.

These particles are predicted by super-symmetry theories (SUSY) to
be thermally produced at the early stages of the universe (Ellis
2001). The best WIMP candidates are neutralinos, but other types of
WIMPs (axions, axinos, gravitinos and Wimpzillas) are not excluded as
an alternative to the constituents of the dark matter component
(Roszkowski 2001).  Nor are charged super-heavy particles with strong
interactions (Albuquerque {\it et.al.} 2000).

It is usually assumed that the WIMPs are a thermal relic of the big
bang. The resulting thermic relic particle should have a mass inferior
to $340 TeV$, the unitary limit (Griest and Kamionkowski 1990).  However,
for SUSY models, this bound is typically 2 orders of magnitude stronger
(Jungman {\it et.al.} 1996).
The neutralino is the best candidate to be produced thermally, while
CDM axinos, gravitinos and Wimpzillas are the best candidates to be
produced by a non-thermal mechanism. For example, in the early
Universe, vacuum fluctuations during or after inflation can produce a
class of very weak super-massive particles with mass in the range of
$10^{12} GeV$ up to $10^{13} GeV$
(Kuzmin \& Tkachev 1999).

Many experiments around the world are currently engaged in searching
for dark matter by assuming that it is dominated by the light
neutralinos, as predicted by the minimal supersymmetric extension of
the Standard Model (Ellis 2001). In particular, the DAMA collaboration
(Bernabei {\it et al.} 2000) has reported evidence of an annual
modulation of recoil energy, which was interpreted as being due to
scattering of some cold dark matter particles with masses between 50
and 100 GeV, and spin-independent cross-sections on a proton
of between
$10^{-42} cm^2$ and $10^{-41} cm^{2}$.

Neutralinos might not constitute all the cold dark matter, but could
be complemented by other particles such as axions and superheavy
relics. Neutralinos are neutral Majorana particles, the lightest SUSY
particles that are massive and stable (Jungman {\it et.al.}
1996). Neutralino properties as dark matter candidates are quite
dependent on the neutralino type, such as the bino, the wino and the
two neutral higgsinos (Rowszkowski 2001).  The neutralino
cross-section for elastic scattering on a nucleus is expected to be
typically very small, roughly $10^{-42}{\rm cm}^2$. This is because
the elastic cross-section is related to the cross-section of
neutralino annihilation in the early Universe which must be only a
fraction of the weak interaction strength in order to have $\Omega_x
h^2\sim 1$ (Bergstr\"om 2000).  In this paper we adopt the neutralino
as our WIMP test particle, unless stated otherwise.

The Sun traps WIMPs in the course of its lifetime, as we discuss
below.  If the Sun were to contain even a minute mass fraction of
WIMPs, there could be a significant influence on its central
thermal structure. Helioseismology provides a means for an
independent test of the validity of this  idea  and can be used to
discriminate between possible candidates for the Dark Matter
({Kaplan {\it et al.}} 1991). Indeed, helioseismology has become a
mature discipline through the detection of more than three
thousand solar acoustic modes with which it has been possible to
constrain the internal structure of the Sun (Gough 1996).  The
large quantity of seismic data obtained by the ground-based
network during the last 10 years, as well as the three seismic
instruments on board the SOHO mission during the last 5 years,
contributed to improve the accuracy of the frequency determination
to a level of 1 part in $10^{5}$ (Bertello {\it et al.} 2000;
Garcia {\it et al.} 2001). This has allowed us to obtain a solar
model for which the equilibrium thermodynamical structure
accurately reproduces the observed seismic data, the so-called
seismic solar model.

In this paper, we discuss how WIMPs can modify the evolution of the
Sun and ultimately the present internal structure, namely in the
nuclear region. In particular, we are interested in WIMPs that
interact with baryonic matter through a weak interaction with a
spin-independent scattering cross-section with values in the range of
the current experiments for the detection of dark matter, such as the
DAMA collaboration, and other ongoing or future experiments.
It follows from our analysis that the possible WIMP candidates
proposed by the DAMA experiment are in disagreement with our
current models.

The next section contains a description of the present status of the
standard solar model, and the current view of the different
difficulties in modelling the observed Sun. In Section~3, we present a
theoretical description of WIMPs and  how they interact with baryonic
matter. In particular, we discuss qualitatively  the type of WIMPs
 capable of modifying the evolution of the Sun.  In Section~4, we
discuss solar models that evolve in halos of dark matter, and we
discuss how those WIMPs can modify the evolution of the Sun. We
conclude with a discussion of our new results, stressing
their pertinence to the current status of the dark matter problem and
Cosmology.

\section{THE PRESENT SUN: THE SOLAR STANDARD MODEL}

The Sun is a unique star for research because its proximity allows
a superb quality of solar data, enabling precision measures of its
luminosity, mass, radius and chemical composition. The Sun
therefore naturally becomes a privileged target for testing
stellar evolution theory. In recent years, different groups around
the world have produced solar models in the framework of classical
stellar evolution, taking into account the best known physics as
well as all the available observational seismic data. This has led
to determination of a well-established model for the Sun, the
so-called standard solar model  (Turck-Chi\'eze \& Lopes 1993;
Christensen-Dalsgaard {\it et al.} 1996; Brun, Turck-Chi\`eze \&
Zahn 1999; Provost, Berthomieu \& Morel 2000; Turck-Chi\'eze,
Nghiem, Couvidat \& Turcotte 2001, Turck-Chi\`ze S. {\it et al.}
2001a; Bahcall, Pinsonneault, Basu 2001), for which the acoustic
modes are in very good agreement with observation. Furthermore,
this model has established considerable consensus among the
different research groups, concerning the predictions of the solar
neutrino fluxes, and has unambiguously helped define the
difference between the theoretical predictions and the
experimental results.

The combined effort between helioseismology and stellar
astrophysics has contributed to strongly constrain the internal
structure of the present Sun.  Indeed, among the different seismic
diagnostics that the solar model is tested against, the radial
profile of the square of the sound speed is known with a precision
as high as $0.3\%$, and in particular in the nuclear region this
precision goes up to $0.2\%$ (see Fig~1). This progress has been
achieved through a detailed description of the microscopic
physics, such as the nuclear reaction rates, the equation of
state, the coefficient of opacities and the gravitational settling
of chemical elements.  More recently, the macroscopic mixing
induced by differential rotation, seems to present a determining
role in the evolution of the star. These physical processes take
place in the thin shear layer, located between the radiative
interior and the convection zone, the so-called tachocline
(Elliott \&  Gough 1999).

The standard stellar evolution of the Sun assumes that the star is
in hydrostatic equilibrium, is spherically symmetric and that the
effects of rotation and magnetic fields are negligible. The
present solar structure and evolution are computed starting from
an initially homogeneous star of a given composition. The solar
models used in this paper have been computed by using the CESAM
code (Morel 1997). The main physical inputs are as follows. We
used the solar composition of Grevesse, Noels \& Sauval (1996),
the OPAL96 tables of Iglesias \& Rogers(1996), the equation of
state of Rogers, Stenson \& Iglesias (1996), the nuclear reaction
rates of Adelberger et al. (1998), the Mitler (1977) prescription
for the treatment of the nuclear screening rates, and the
microscopic diffusion coefficients suggested by Michaud \& Proffit
(1993). The present structure of the Sun is obtained by evolving a
initial star from the pre-main sequence, around 0.05 Gyr from the
ZAMS, until its present age, 4.6 Gyr. The present solar model is
obtained by choosing the initial Helium content, as well as the
mixing length parameter of convection that best predicts the
present solar luminosity and radius. The helioseismolgy requires
that the present structure of the Sun, including the global
quantities, should be determined with a  very high precision. In
particular, the calibration of the solar radius is done with a
precision of $10^{-5}$.

\section{THE EVOLUTION OF THE SUN IN THE PRESENCE OF  DARK MATTER}

In order to explain the rotation of spiral galaxies, it is
necessary to assume that these are immersed in a halo of dark
matter. In the present model, we assume the halo of the Milky Way
to be an isothermal gas of WIMPs, the distribution of their
velocities being Maxwellian with a mean velocity $\bar{v}\sim
240\; Km\;s^{-1}$. Furthermore, we assume that the local dark
matter density near the Sun is about $0.3 \; GeV\;cm^{-3}$
(Jungman {\it et.al.} 1996).  Typically, the local density by
number is of the order of $3\;10^{-2}\; cm^{-3}$ for WIMPs with
$m_{\rm x}\sim 10 {\rm GeV}$, and $3\;10^{-3}\;cm^{-3}$ for WIMPs
of $m_{\rm x}\sim 100 \;{\rm GeV}$.

\subsection{The WIMP scattering cross-section}

Any WIMP that crosses the interior of the Sun may interact with a
nucleus and lose enough energy to be trapped by the star's
gravitational potential well. The WIMP interactions with the
baryonic matter will depend upon their mass, $m_{\rm x}$ and on
their scattering cross-section on the nucleons, $\sigma_p$.  The
elastic scattering cross-section of relic WIMPs scattering off a
nucleus in the solar interior depends on the individual
cross-sections of the WIMPs scattering off of quarks and gluons.
For non-relativistic Majorana particles, such as the neutralino,
these can be divided into two separate types.  The coherent part
described by an effective scalar coupling between the WIMP and the
nucleus is proportional to the number of nucleons in the nucleus.
The incoherent component of WIMP-nucleus cross section results
from an axial current interaction of a WIMP with the constituent
quarks, and couples the spin of the WIMP to the total spin of the
nucleus. These two types of cross-sections have been computed
exactly (Goodman \& Witten 1985, Jungman {\it et.al.} 1996, and
references therein), and in the case of scalar interaction it
 strongly depends on the particle physics model used. At this
stage, it is worth noticing that in the case of scalar (or
coherent) interactions, the total scattering cross-section is
proportional to the mass of the nucleus, $ \sigma^{sc}\sim \mu_i^2
$, where $\mu_i=m_x m_i/(m_x+m_i)$ is the reduced mass, and in the
case of axial (or incoherent) interactions, it can be shown that
the cross-section, $ \sigma^{ax}\sim \mu_p^2 $, depends on the
spin of the nucleus as $J(J+1)$.  In the case of scalar
interactions, because of the constructive interference between all
nucleons in the nucleus, the cross-section rises rapidly with the
atomic number and, as a consequence, WIMPs scatter much more
efficiently on the heavier nuclei. Inside the Sun, elements
heavier than hydrogen contribute substantially to the energy
transfer and to the capture rate.
Nevertheless, the Sun's evolution on the main sequence is strongly
dominated by the production of helium rather than heavier chemical
elements. In such a scenario, the helium nucleus will be the
dominant source of scalar scattering, mainly due to its high
abundance, given that heavier nuclei such as iron, nitrogen,
carbon and oxygen are sub-dominant species. However, in more
detailed studies on the capture of WIMPs by the Sun, the
contribution of heavier elements seems to be more significant than
was previously thought (Gould 1987). In fact, the scattering on
these species will increase the total number of WIMPs accreted by
the Sun leading to a more significant change on the structure of
the solar core. Consequently, by choosing to consider the
scattering only for helium, we make a conservative approach to
study the impact of WIMPs on stellar evolution and
Helioseismology.

The essential particle physics parameters which enter the dark matter
problem are the mass of the WIMP $m_x$ and the elastic scattering
cross-section, $\sigma_p$, of the various nuclei which constitute
the solar material, and $\langle\sigma_a v\rangle$ the thermal average
of the annihilation cross-section times the relative velocity of the
WIMPs at freeze-out. The annihilation cross section can be calculated
for each model (Krauss {\it et.al.} 1985)
however, since we lack certain theoretical guidance, we
decided to concentrate on a typical simple case for WIMP-nucleus
scalar scattering, as previously discussed. We relate the
cross-section of the WIMP on nuclei to the WIMP-proton cross section
$\sigma_p$ taken as a free parameter.

\begin{figure}
\centerline{\psfig{file=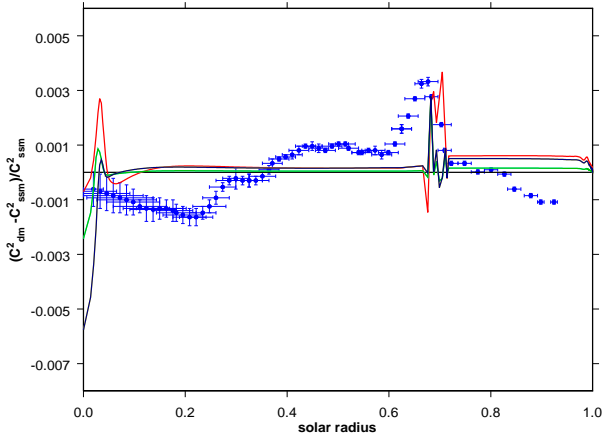,width=9.3cm,height=7.0cm}}
\caption{ The relative differences between the square of the sound
speed of the standard solar model and solar models, evolving
within a halo of WIMPs.  The continuous curves correspond to
models with WIMPs with masses of 60 GeV, annihilation
cross-section of $10^{-32}cm^3/s$, and scalar scattering
cross-section of $10^{-38} cm^2$ (red curve), $10^{-36}cm^2$
(green curve) and $10^{-40}\;cm^2$ (dark blue curve).
The peculiar form of these profiles occur as a consequence of the
redistribution of energy by WIMPs inside the star. In the core
($r<0.2R_\odot$) the transport of energy by WIMPs decreases the
temperature and the molecular weight leading occasionally to a
reduction of the central sound speed. Reversely, in the outer
layers, the energetic balance imposed by the calibration of the
Sun, leads to an increase of the sound speed over all the solar
envelope. The strong signature on the base of the convection zone
($r\approx 0.7R_\odot$) is related with the effect on stellar
evolution produced by the shear layer, the so-called tachocline.
However, this effect should be carefully considered, once that it
still remains an important source of incertitude.
The curve with error bars represents the relative differences
between the squared sound speed in the Sun (as inverted from solar
seismic data) and in a standard solar model (Turck-Chi\'eze,
Nghiem, Couvidat \& Turcotte 2001; Kosovichev {\it et al.}
1999;1997). The horizontal bars show the spatial resolution and
the vertical bars are error estimates.}\label{fig:1}
\end{figure}

\subsection{The Capture and Annihilation Rates of WIMPs in the Sun}

The abundance of WIMPs in the Sun depends on the WIMPs accumulated
in the Sun by capture from the Galactic halo and is depleted by
annihilation. If $N_x$ is the number of WIMPs in the Sun, then the
differential equation governing the time evolution of $N_x$ is
(Gould {\it et.al} 1987)
\begin{eqnarray}
\frac{dN_x}{dt}=\Gamma_c- C_a\;N_x^2,
\end{eqnarray}
where $\Gamma_c$ is the rate of accretion of WIMPs onto the Sun.  The
determination of $\Gamma_c$ depends on the nature of the particle. If
the halo density of WIMPs remains constant in time, $\Gamma_c$ is
time-independent. The second term accounts for depletion of WIMPs, and
is twice the annihilation rate in the Sun, $\Gamma_a=1/2\;C_a N_x^2$.
The quantity $C_a$ depends on the WIMP total annihilation cross
section times the relative velocity in the non-relativistic limit,
$\langle\sigma_a v\rangle$, and the distribution of WIMPs in the
Sun. The total annihilation cross-section of the new particle should
possibly have a value that is consistent with the cosmological density
of dark matter as specified, $\langle\sigma_a v\rangle \approx 3 \,
10^{-26} cm^3 s^{-1} \Omega_x^{-1} h^{-2} $. It follows that, not
knowing the initial content of relic WIMPs, $\Omega_x$, we will
consider different scenarios, such that $\langle\sigma_a v\rangle <
10^{-26} cm^3 s^{-1} $ (Jungman {\it et.al.} 1996).

Solving the previous first-order differential equation, the total
number of WIMPs at  time $t$, is given by
\begin{eqnarray}
N_x(t)=\Gamma_c\;\tau\;\tanh{\left(\frac{t}{\tau}\right)},
\end{eqnarray}
and the annihilation rate is given by
\begin{eqnarray}
\Gamma_a(t)=\frac{1}{2}\;\Gamma_c\;\tanh^2{\left(\frac{t}{\tau}\right)},
\end{eqnarray}
where $\tau=1/\sqrt{\Gamma_c \; C_a}$ is the time scale for capture
and annihilation of WIMPs to equilibrate.
As the age of the Sun is much larger than the equilibrium
time-scale, $t_\odot/\tau \gg 1$,  equilibrium is reached very
rapidly, then $\Gamma_a=1/2\; \Gamma_c$.
 It follows that
\begin{eqnarray}
\frac{t_\odot}{\tau}=322
\left(\frac{\Gamma_c}{s^{-1}}\right)^{1/2}\left(\frac{\langle\sigma_a
v \rangle}{cm^3
s^{-1}}\right)^{1/2}\left(\frac{m_x}{10GeV}\right)^{3/4},
\label{eq-ttau}
\end{eqnarray}
where $t_\odot$ is the present age of the Sun
(Jungman {\it et.al.} 1996).

The rate of accretion of WIMPs in the Sun was first calculated by
Press and Spergel (1985).  Even though the detailed computation of
the accretion can be quite elaborate, the basic idea is simple
(Srednicki {\it et.al.} 1987); if the WIMP elastically scatters
from a nucleus with a velocity smaller than the escape velocity,
$v_{esc}=\sqrt{2GM_\odot/R_\odot}$, then the WIMPs will be
captured. The WIMPs have a typical velocity of the order of $300\;
km/s$,
 which is smaller than the escape velocity at the surface of the
 Sun $618\; km/s$, consequently, they are trapped quite efficiently in
 the solar interior. If we consider that all the WIMPs that hit the Sun
 lose enough energy to be trapped, the trapping rate is given by the
 product of the surface area of the Sun, about $6.1\;10^{22}\; cm^2$,
 and the flux of WIMPs, of about $n_{\rm x}\;\bar{v}$. It reads $
 \Gamma_{\rm c}=4\pi R_\odot^2 n_{\rm x} \bar{v}$.  Gravitational
 focussing effects enhance the previous trapping rate by
 $2GM_\odot/\bar{v}^2$. This occurs because of the enlargement of the
 Sun's effective area due to the gravitational well and the requirement
 that the impact parameter be less than $R_\odot$.  This yields a
 trapping rate of $\Gamma_{\rm c}=\pi(2GM_\odot R_\odot) n_{\rm
 x}/\bar{v}$. Bouquet and Salati (1989) have generalized and refined
 the capture rate for main-sequence stars (Salati \& Silk 1989)
\begin{eqnarray}
\Gamma_c=10^{32}\; {\rm s^{-1}} \left(\frac{\rho_{\rm
x}}{1\;M_\odot {\rm pc}^{-3}}\right) \left(\frac{m_p}{m_x}\right)
\left(\frac{300\;{\rm km}{\rm s}^{-1}}{\bar{v}_x}\right) \nonumber
\\
\left[1+0.16\left(\frac{\bar{v}_x}{300\;{\rm km}{\rm
s}^{-1}}\right)^2 \left(\frac{M_\odot}{M}\right)
\left(\frac{R}{R_\odot}\right) \right] \nonumber
\\
\left(\frac{M}{M_\odot}\right) \left(\frac{R}{R_\odot}\right) {\rm
min}\left[1,\frac{\sigma_p}{\sigma_\odot}\frac{M}{M_\odot}
\left(\frac{R_\odot}{R}\right)^2\right].
\end{eqnarray}
The total number of WIMPs accreted, $N_x$, can be computed taking into
account that $t_\odot/\tau >> 1$, and is
\begin{eqnarray}
N_x\simeq \Gamma_c\tau.
\end{eqnarray}
The total number of WIMPs at present strongly depends on the ratio $
\sqrt{\langle\sigma_a v\rangle/\Gamma_c}$. If we consider $
\langle\sigma_a v\rangle$ to be constant, the capture is weak and
proportional to the scattering cross-section and $N_x$ increases with
$\Gamma_c$, for the case of low scattering cross-sections. On the
contrary, for large scattering cross-sections, any WIMP which enters
the star is captured, the captured flux saturates at the level of the
incoming flux and the system reaches an equilibrium.
Note that evaporation is unimportant for WIMPs
in the mass range considered here, namely  $m_x>10$ GeV (Gould 1987).

This concentration of WIMPs in the solar core has a marginal effect on
the evolution of the Sun. Luckily, the precision  presently obtained by
seismic diagnostics allows  to determine in certain cases the effect of
WIMPs on the solar core. In fact, a more accurate expression for the
capture should take into account the way that WIMPs scatter from the
nucleus through scalar interactions and/or axial-vector interactions,
as well as second-order effects (Jungman {\it et.al.} 1996). However,
we choose to focus on a simplified picture in this first approach to
the problem.

\subsection{The energy transport by WIMPs in the solar interior}

The energy transport by WIMPs in the solar interior is governed by
three natural length scales: the mean free path of the WIMP, $l_x$,
the inverse of the logarithmic temperature gradient, $\nabla \ln{T}$,
and a typical geometric dimension of the system, such as the WIMP
scale height, $r_x$. Bouquet and Salati (1989) showed that the WIMP
distribution is approximatively gaussian and therefore
\begin{eqnarray}
n_x(r)=n_0\;\exp{\left[-\frac{r^2}{r_x^2}\right]}
\end{eqnarray}
where $n_0=N_x/\pi\sqrt{\pi} r_x^3$, and $r_x^2 = 3 T_c/(2 \pi m_x G \rho)$
which typically is $r_x \approx 10^{-2} \, R_\odot \, m_{100}^{-1/2}$
with $m_{100}$ the WIMP mass in units $100$ GeV.

There are two extreme regimes for the energy transport by WIMPs,

characterized by the values of the Knudsen number, $K$, defined as the
ratio $l_x/r_x$. The Knudsen number strongly depends on the scattering
cross-sections of the WIMPs on nuclei, and typically goes like $K
\approx 30 \, m_{100}^{3/2} / \sigma_{36}$ where $\sigma_{36}$ is the
scattering cross section measured in units $10^{-36}$ cm$^2$.

In the case of small cross-sections, the WIMP mean free path is larger
than the scale length $r_x$, $K > 1$, and two successive collisions
are widely separated. This is the so-called large Knudsen number
regime (or Knudsen regime). The transport of energy is non-local. Thus
the WIMPs undergo few interactions on each orbit and therefore
are not in thermal equilibrium with the nuclei. In such a case, it is
convenient to define the average temperature of the WIMP core, $T_x$,
as being the temperature of the star at $r_x$. At the center of the
star, the WIMP temperature (or the averaged kinetic energy) is lower
than the temperature of the nuclei, and in such conditions the net
effect of the WIMP-nucleus collisions is to transfer energy from
nuclear matter to WIMPs.  Conversely, in the outskirts of the WIMP
core, in certain cases the WIMPs give back the energy gained in the
center of the star.  This process is a very efficient mechanism at
evacuating the energy from the nuclear region. In such a regime no
precise analytic result exists. Nevertheless, an approximative
solution has been found by Spergel and Press (1985). The luminosity
carried by WIMPs can then be evaluated by
\begin{eqnarray}
L_{\rm lp}(r)=32\sqrt{2\pi}\;\frac{\sigma_p\mu_p}{m_x+m_p}
\int_0^r n_p(r) \bar{T}(r) n_x(r)r^2dr
\end{eqnarray}
where $\bar{T}$ reads
\begin{eqnarray}
\bar{T}(r)=\sqrt{\frac{m_pT_x+m_x T(r)}{m_xm_p}}\left[T(r)-T_x\right].
\end{eqnarray}

In the opposite limit, $K < 1$, the transport of energy is local and
the energy transport proceeds by conduction, allowing for a much
simpler treatment (Gilliland {\it et.al.} 1986). In this regime, the
WIMP mean free path is much smaller than the dimension of the region
where they are trapped. The WIMPs interact sufficiently often to be in
local thermal equilibrium with the nuclei and therefore it is possible
to define a temperature of the WIMP, which is equal to the temperature
of the nuclei.
\begin{eqnarray}
L_{\rm sp}(r)=-4\pi r^2 \;n(r)\;\sqrt{\frac{T(r)}{m_x}} l_{sp}(r)\;
\nabla T(r)
\end{eqnarray}
where $l_{sp}(r)$ is the local mean free path of the WIMP,
it reads
\begin{eqnarray}
l_{\rm
sp}(r)=\left[\frac{\rho(r)}{m_u}\sum\sigma_i\frac{X_i(r)}{A_i}\right]^{-1}.
\end{eqnarray}
The range of WIMP masses allowed by the cosmological model is quite
large, even if we restrain ourselves to the WIMPs that are produced
in a thermal scenario. Consequently, depending also on their
scattering cross-section, the WIMPs can transport energy, not only in
the conductive region and in the Knudsen regime, but also in an
intermediate case. We have chosen to define the intermediate case as
the one where the transport of energy is determined by the
interpolation formula (Gilliland {\it et.al.} 1986)
\begin{eqnarray}
L_x(r)=\frac{L_{lp}(r)L_{sp}(r)}{L_{lp}(r)+L_{sp}(r)}.
\end{eqnarray}
This formula appropriately converges to the right approximation in
each of the two regimes, accordingly to the value of $K$. The errors
introduced are within the precision of our computation.

\section{DISCUSSION AND CONCLUSIONS}

\begin{figure}
\centerline{\psfig{file=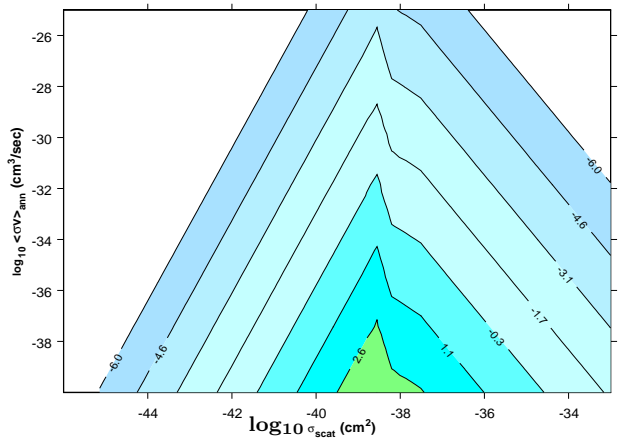,width=9.3cm,height=7.0cm}}
\vspace{-0.8cm} \centerline{{\hspace{-1.6cm}\small \sf \bf
log$_{\bf 10}$}}\vspace{-0.1cm}
 \caption{This figure presents the relative variation of the
luminosity from the core of the Sun, due to the concentration of
WIMPs in the solar core, as a function of scalar scattering cross
section and annihilation cross section.  The computation was made
for a WIMP mass of 111 GeV.  The iso-curves represent the decimal
logarithm of the ratio of the luminosity of WIMPs against the
Sun's luminosity in the inner core of 5\% of the solar radius. The
computation was made using the structure of the present Sun
($t_\odot\sim 4.6 \; Gyr$) which was obtained by using an
evolution code that assumes the evolution of the Sun under a
standard evolution.
In the solar models with  scattering cross sections lower than
$10^{-38}\; cm^2$, the increase of the luminosity of WIMPs with
the scattering cross section is due to the increase on the number
of particles trapped in the Sun's core. Alternatively, for the
models with scattering cross sections above $10^{-38}\;cm^2$, the
population of WIMPs inside the star is in equilibrium, and the
increase of the scattering cross section leads to a less efficient
energy transport  by WIMPs,
reducing the WIMP luminosity on the solar core.}
\label{fig:2}
\end{figure}
The thermodynamical structure in the interior of the Sun, namely
in the nuclear region, is presently known with a precision of much
less than a few per cent. This level of accuracy in constraining
the solar interior has been achieved by a systematic study of the
differences between the acoustic spectrum obtained from
helioseismology experiments and the theoretical spectrum.
Presently, this difference is less than $10 \mu Hz$  for almost
all of the 3000 modes that probe the interior of the Sun (Gough
1996).
This level of precision in the description of the solar core
allows us to discriminate physical processes that could not be
discussed otherwise, in particular those that present a peculiar
behaviour, as  seems to be the case for WIMPs trapped in the solar
core. In Fig.~1, we compare the square of the sound speed,
$c_s^2$, of different solar models evolving within the presence of
a halo of WIMPs and the solar standard model (Brun, Turck-Chi\`eze
\& Morel 1998). The changes induced by the presence of WIMPs are
concentrated in the inner core within 10\% of the solar radius,
typically seen in the profiles of the temperature, density and
molecular weight.

Indeed, the WIMPs are thermalized within the solar core and are on
Keplerian orbits around the solar center, interacting through
elastic scattering with the solar nuclei, such as helium, and thereby
providing an alternative mechanism of energy transport other than
radiation. The result is a nearly flat temperature distribution,
leading to an isothermal core. Consequently, the central
temperature is reduced. This reduction of temperature has two main
consequences: since central pressure support must be maintained,
due to the hydrostatic equilibrium,  the central density is
increased in the solar models with WIMPs, and since less hydrogen
is burnt at the centre of the Sun, the central helium abundance
and the central molecular weight are smaller than in standard
solar models. The increase of the central density and hydrogen
partially offset the effect of lowering the central temperature in
the central production of energy. In fact, this is the reason why
minor changes are required to the initial helium abundance and the
mixing-length parameter in order to produce a solar model of the
Sun with the observed luminosity and solar radius. This readily
leads to a balance between the temperature, and the molecular
weight  in the core, leading to the peculiar profile of the square
of the sound speed, $c_s^2\propto T/\mu$. This seems to be the
case for most of the solar models within WIMP halos. The balance
between the temperature and molecular weight is critical for the
profile in the center of the star, leading to some of the profiles
presented in Fig.~1.
In the same figure, we display the inversion of the sound speed
obtained from the data of the three seismic experiments on board
the SOHO satellite. It follows from our analysis that the presence
of WIMPs in the solar core produces changes in the solar sound
speed of the same order of magnitude as the difference between the
sound speed of the standard solar model and the inverted sound
speed. It is important to remark that the inversion of the sound
speed still presents some uncertainty in the central region due to
the lack of seismic data, mainly due to the small number of
acoustic modes that reach the nuclear region. Furthermore, the
inversions are not very reliable at the surface, above $98 \%$ of
the solar radius, due to a poor description of the interaction of
acoustic waves with the radiation field and the turbulent
convection, namely, in the superadiabatic region (Lopes \& Gough
2001). However, we can establish with certainty that the
difference between the inverted sound speed and the theoretical
sound speed is known with a precision of $0.3\%$ in the solar
interior, within $95\% $ of the solar radius. The evolution of the
Sun in a halo of WIMPs will increase the evacuation of energy from
the solar core. The WIMPs will work as a 'cold bridge' between the
core and the more external layers of the Sun. The magnitude of the
effect is proportional to the total number of WIMPs concentrated
in the solar core.
Nevertheless, even if some systematic effect is present in the
inversion of the sound speed, the presence of WIMPs in the solar
core leads to a quite different nucleosynthesis history from  the solar
standard model case, and from that to a peculiar radial profile of
the sound speed. In this way, the effect of WIMPs in the solar
core can be inferred on the basis of seismic diagnostics, such as
the inversion of the square of the sound speed, among other possible
techniques. The  proposed method constitutes a new way to
disentangle the contribution of  different non-baryonic
particles to the dark matter.

The luminosity in the core of the Sun is presently known with a
precision of one part in $10^{-3}$. In the coming years, it is
very likely that the  new seismic data available from the SOHO
experiments, will allow us to obtain a seismic model of the Sun
with an accuracy of $10^{-5}$.  In such conditions, the Sun can
and should be used as an excellent probe for dark matter in our
own galaxy. In Figs.~2 and~3, we compute the ratio of the WIMP
luminosity against the Sun's luminosity produced in the inner core
of $5\%$ of the solar radius. A significative region of the
$\sigma_{scat}-\langle\sigma_{a}v\rangle$ plot shows changes in
the solar luminosity of the order of $10^{-3}$. This order of
magnitude on the luminosity produced in the solar core can be
tested through seismological data. In particular, we are
interested in the lighter WIMPs, $m_x< 100GeV$, and the smallest
scattering cross-section, $10^{-45}\; cm^2 $ up to
$10^{-40}\;cm^2$. This range of parameters are presently being
tested by the DAMA and CDMS experiments, among others, and are
also well within the range of the parameters of future
helioseismological experiments (see Fig.~3).   It is interesting to note, that
 the X-ray Quantum Calorimeter (XQC) experiment (D. McCammon {\it
 et.al.} 2001) may exclude scattering cross sections bigger than
 about $10^{-29}$ cm$^2$ for the mass range considered here,
 relevant for   strongly interacting massive particles recently
 hypothesised as an alternative dark matter candidate to WIMPs and
 that  penetrate neither subterranean
laboratories nor the solar core  (Wandelt {\it
et.al.} 2000),   hence being complementary to helioseismology
in potentially excluding overlapping regions in parameter space.

\begin{figure}
\centerline{\psfig{file=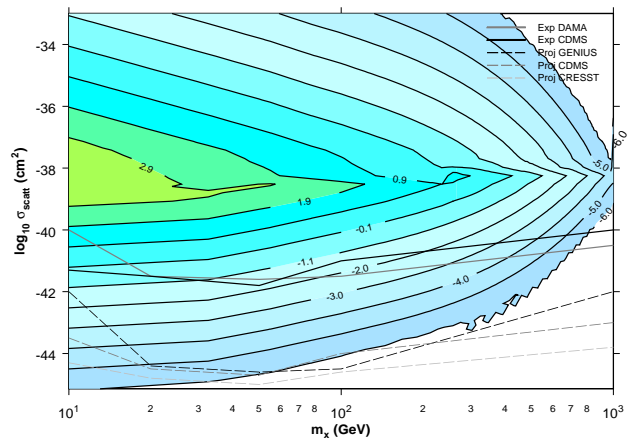,width=9.3cm,height=7.0cm}}
\caption{This figure illustrates the relative variation of the
core luminosity of the Sun, produced in the solar interior by the
presence of WIMPs, as a function of WIMP mass and scattering
cross-section. Annihilation cross section is $\langle\sigma_a
v\rangle = 10^{-36}$ cm$^2$.  The representation is the same as
the one presented in Fig.~2. The solid curves represent the
current experimental bounds placed by DAMA (Bernabei {\it et al.}
2000; black grey solid line) and CDMS (Schnee 1999; black solid
line), and the future projects of CDMS (black dashed line), GENIUS
(Baudis 1998; black gray dashed line) and CRESST (Bravin et al.
19999; light grey dashed line).
An important decrease in WIMP luminosity occurs for solar models
with WIMPs with larger masses. This effect is produced by the
reduction of the WIMP core, so that heavier  WIMPs are more and
more confined to the center of the star.}
 \label{fig:3}
\end{figure}

If we believe in the simple model presented in this paper,
 DAMA candidate WIMPs with
masses of 60 GeV and annihilation cross-section of the order of
$10^{-32}cm^3/s$ cannot exist (see Fig.~1), otherwise their effect in
the solar core should already have been identified by seismic
diagnostics. However, we stress that in order to determine with
certainty the range of masses and cross-sections for WIMPs that is in
disagreement with the helioseismological results, a more careful
analysis of the different regimes of energy transport by WIMPs should
be done. Furthermore, the microscopic physics of this region of the
star is not fully established, as there remains
uncertainty in certain
nuclear reaction rates, such as the p+p reaction and the dynamical
screening in other nuclear interactions such as $^3He+^4He$ and
$^4He+^4He$.

An important diagnostic of the core can be obtained from the
seismology of gravity modes. Indeed, it is the low-degree internal
gravity modes that are the most sensitive to the conditions in the
core, the region where substantial deviations from the so-called
standard solar model might occur. Solar models evolving in the
presence of dark matter have a g-mode period spacing that is
drastically different from that of other solar models, mainly due
to the peculiar distribution of density on the nuclear region that
occurs as a consequence of the energy transport enhanced by the
WIMPs. Therefore, g-mode observations could provide a sensitive
test to the radial distribution of density in the solar core, and
ultimately to the presence of dark matter in the solar
neighborhood. The observation of gravity modes by SOHO seismic
experiments, such as GOLF (Turck-Chi\'eze {\it et al.} 2001b),
could ultimately strongly constrain the physics of the solar core.

We have in this paper focused on a slightly simplified version of
WIMP energy transport.  There are several improvements which can
be included in the analysis (Lopes {\it et al.} 2001), e.g. in the
Knudsen regime the deviation from isotropy leads to a radius
dependent luminosity suppression (Gould \& Raffelt 1990b); in the
conductive regime, the mass dependence of the thermal conductivity
gives an additional factor $\sim 2 \sqrt{m_{100}}$ for scattering
on helium (Gould \& Raffelt 1990a).  Also inclusion of scattering
off other light elements might affect the results slightly.
Finally, the transition region between non-local and conductive
energy transport could be expressed in terms of the Knudsen number
(Dearborn {\it et al.} 1991; Kaplan {\it et al.} 1991) instead of
the intuitive interpolation formula used here, which would be
useful for a detailed investigation of the transition region.

In conclusion, we have identified the range of WIMP masses, scattering
cross-sections and annihilation cross-sections which reduce the
luminosity in the core of the Sun. These effects on the solar
structure are now within the range of effects capable of being probed
by the diagnostic capabilities of helioseismology. We did not concern
ourselves with particular particle physics models, but considered a
generic case, for WIMPs which have a large range of masses, scattering
cross-sections and annihilation cross-sections.  The effect of WIMPs
in the core of the Sun is of the same order of magnitude as the
microscopic physics and dynamical processes that are now being
discussed in the framework of stellar evolution in the light of the
most recent results of helioseismology.  Studies of the sun and of
cosmology may have much to gain from each other.

\section*{Acknowledgments}

 The authors thanks P. Morel for using the CESAM code and S.
Turck-Chi\`eze and R. Garcia for stimulating discussions and
helioseismic analysis.
The authors would like also to thank the referee D. Jungman for
his valuable suggestions that lead to improve the original
manuscript.
 SHH is supported by a
Marie Curie Fellowship of the European Community under the
contract HPMFCT-2000-00607. IPL is grateful for support by a grant
from Funda\c c\~ao para a Ci\^encia e Tecnologia.



\begin{thebibliography}{99}

\bibitem{Adelberger:1998}
Adelberger, E., et al. 1998, Rev. Mod. Phys., 70, 1265

\bibitem{Albuquerque:2000rk}
Albuquerque I. F., Hui L. and Kolb E. W., 2000
hep-ph/0009017.


\bibitem{Bahcall:2001a}
Brun  A. S., Turck-Chi\`eze S., Zahn J. P. 1999, ApJ, {\bf 537},
L143.

\bibitem{Bahcall:2001}
Bahcall J. N., Pinsonneault M. H., Basu S. 2001, ApJ, {\bf 555},
990.

\bibitem{Bergstrom:2000pn}
Bergstrom L., 2000
Rept.\ Prog.\ Phys.\  {\bf 63}, 793 [hep-ph/0002126].




\bibitem{Bernabei91}
 Bernabei R. et al. [DAMA Collab.] 1997,
 Phys. Lett. B389, 757.

\bibitem{Bernabei:2000qi}
Bernabei  R. {\it et al.}  [DAMA Collaboration] 2000,
Phys.\ Lett.\ B {\bf 480}, 23.



\bibitem{Elliott:1999f}
 Bertello et al. 2000 ApJ, {\bf 537}, L143.



\bibitem{Bouquet:1987ps}
Bouquet A. and Salati P., 1989, Astron. Astrophys. {\bf
217},270-282.


\bibitem{Bravin99}
Bravin M. et al. (CRESST Collab.) 1999, hep-ex/9904005.



\bibitem{Elliott:1999a}
Brun, Turck-Chi\`eze and Morel 1998 ApJ, {\bf 506}, 913.




\bibitem{Elliott:1999b}
Brun, Turck-Chi\`eze and Zahn 1999 ApJ, {\bf 368}, 626.



\bibitem{Baudis98}
Baudis L. et al. (GENIUS Collab.) 1998, Phys. Rep. 307, 301.



\bibitem{Elliott:1999c}
Dearborn D., Griest K., Raffelt G. 1991 ApJ, {\bf 368}, 626.



\bibitem{Elliott:1999d}
Elliott J. R., Gough D. O., 1999 ApJ, {\bf 516}, 475.


\bibitem{Ellis:2001}
Ellis J. 2001, Nucl. Phys. Proc. Suppl. 101, 205-216
[hep-ph/0103288].


\bibitem{Faulkner:1985rm}
Faulkner J. and Gilliland R. L. 1985,
Astrophys.\ J.\  {\bf 299}, 994.

\bibitem{gilliland:1986}
R. L. Gilliland,  J. Faulkner W. H. Press and D. N. Spergel,
Astrophys.\ J.\  {\bf 306} (1986) 703.


\bibitem{Garcia:2001}
 Garcia R. et al., 2001, Solar Physics, {\bf 200}, 361.


\bibitem{Goodman:1985dc}
Goodman M. W. and Witten E. 1985,
Phys.\ Rev.\ D {\bf 31}, 3059.





\bibitem{Gough:1996}
Gough D. O. et al., 1996, Sci, 272, 1296


\bibitem{Gould:1990ez}
Gould A. and Raffelt G. 1990,
Astrophys.\ J.\  {\bf 352}, 669.

\bibitem{Gould:1987ju}
Gould A. 1987,
Astrophys.\ J.\  {\bf 321}, 560.


\bibitem{Gould:1990hm}
Gould  A. and  Raffelt G. 1990,
Astrophys.\ J.\  {\bf 352}, 654.

\bibitem{Grevesse:1996}
Grevesse, N., Noels, A., \& Sauval, A. J. 1996, in ASP Conf. Ser.
99, Cosmic Abundances, ed. S. S. Holt \& G. Sonneborn (San
Francisco: ASP).

\bibitem{Griest:1990wd}
Griest K. and Kamionkowski M. 1990,
Phys.\ Rev.\ Lett.\  {\bf 64}, 615.

\bibitem{Iglesias:1996}
Iglesias, C., \& Rogers, F. J. 1996, ApJ, 464, 943.

\bibitem{Jungman:1996df}
Jungman G., Kamionkowski M. and Griest K. 1996,
Phys.\ Rept.\  {\bf 267}, 195 [hep-ph/9506380].

\bibitem{Kamionkowski:1995dp}
Kamionkowski M., Griest K., Jungman G. and Sadoulet B. 1995,
 detection of supersymmetric dark matter,''
Phys.\ Rev.\ Lett.\  {\bf 74}, 5174 [hep-ph/9412213].

\bibitem{Kaplan:1991vf}
Kaplan J., Martin F. de Volnay, Tao C. and Turck-Chi\'eze S. 1991,
Astrophys.\ J.\  {\bf 378}, 315.

\bibitem{krauss:1985}
Krauss L. M., Freese K., Spergel D. M., Press W. H. 1985,
Astrophys.\ J.\  {\bf 299}, 1001.


\bibitem{Kosovichev:1999}
Kosovichev, A. G., 1999, J. Comp. Appl. Math, {\bf 109}, 1.


\bibitem{Kosovichev:1997}
Kosovichev, A. G. {\it et al.}, 1997, Solar Phys, {\bf 170}, 43.

\bibitem{Kuzmin:1999}
Kuzmin V. A., Tkachev I. I. 1998, JETP Lett. 68, 271-275
[hep-ph/9802304].

\bibitem{Lee:1977ua}
Lee B. W. and Weinberg S. 1977,
Phys.\ Rev.\ Lett.\  {\bf 39}, 165.



\bibitem{lopesgough}
Lopes I. P., Gough D. O. 2001, MNRAS {\bf 322}, 473.


\bibitem{lopesetal}
Lopes I. P., Hansen S. H.,  Silk J., 2001 (in preparation).

\bibitem{Mccammon:2001}
McCammon D. {\it et.al.} 2001, in preparation.


\bibitem{Michaud:1993}
Michaud, G., \& Proffitt, C. R. 1993, in IAU Colloq. 137, Inside
the Stars, ed. A. Baglin \& W. W. Weiss (ASP Conf. Ser. 40; San
Francisco: ASP)

\bibitem{Mitler:1997}
 Mitler, H. E. 1977, ApJ,
212, 513

\bibitem{Morel:1997}
Morel P. 1997, A\& AS, 124

\bibitem{Press:1985ug}
Press  W. H. and Spergel D. N.,
Astrophys.\ J.\  {\bf 296} (1985) 679.

\bibitem{provost:2000}
Provost, J.; Berthomieu, G.; Morel, P. Astronomy and Astrophysics
{\bf 353} (2000) 775.


\bibitem{Rogers:1996}
Rogers, F. J., Swenson, J., \& Iglesias, C. 1996, ApJ, 456, 902

\bibitem{Roszkowski:2001yq}
Roszkowski L.,
hep-ph/0102327.

\bibitem{Salati:1987ja}
Salati P. and Silk J., Astrophys.\ J.\  {\bf 338} (1989) 24.



\bibitem{Schnee99}
Schnee R. [for the CDMS Collab.], talk presented at Inner
Space/Outer Space II," Fermilab, May 1999.

\bibitem{Spergel:1985}
Spergel D. N., and Press W. H. 1985, Astrophys.\ J.\  {\bf 294},
663.

\bibitem{Srednicki:1987vj}
Srednicki M., Olive K. A. and Silk J. 1987,
Nucl.\ Phys.\ B {\bf 279}, 804.

\bibitem{Steigman:1978vs}
Steigman G., Sarazin C. L., Quintana H. and Faulkner J. 1978,
{\it  Astron. J. 83, 050-1061. In *Srednicki, M.A. (ed.): Particle
physics and cosmology* 207-218}.

\bibitem{TCL:1993a}
Turck-Chi\'eze S., Lopes I., 1993, Astrophys. J., 408, 347

\bibitem{TCL:1993b}
Turck-Chi\`eze S., Nghiem P., Couvidat S. \& Turcotte S., 2001,
Sol. Phys., {\bf 200}, 323.



\bibitem{turckchieze:2001}
Turck-Chi\`ze S. {\it et al.}  2001a, Astrophys. J., {\bf 555}
L69.

\bibitem{TCLa:2001}
 Turck-Chi\`eze S.,  Garcia R. A.,  Couvidat S.,
Ulrich R. K. , Bertello L., Varadi F., Kosovichev A. G.,
 Berthomieu G.,
Provost J.,   Brun A. S.,  Lopes I. P.,  Robillot J. M.,  Rocca
Cortes T., 2001b, to be submitted to the Astrophysical Journal.


\bibitem{Wandelt:2000ad}
Wandelt B. D., Dave R., Farrar G. R., McGuire P. C., Spergel D. N.
and Steinhardt P. J.,
astro-ph/0006344.

\end{thebibliography}
\end{document}